\documentclass[10pt,preprintnumbers,showpacs]{revtex4}
\usepackage{graphicx,epsfig,axodraw}

\def\xslash#1{{\rlap{$#1$}/}}
\def\half{\frac{1}{2}}
\def\beq{\begin{equation}}
\def\eeq{\end{equation}}
\def\beqa{\begin{eqnarray}}
\def\eeqa{\end{eqnarray}}
\def\iar{\begin{array}{l}}
\def\ear{\end{array}}

\begin{document}

\title{Fermion field renormalization prescriptions}
\author{Yong Zhou}
\affiliation{Institute of High Energy Physics, Academia Sinica, P.O. Box 918(4), Beijing 100049, China, Email: zhouy@mail.ihep.ac.cn}

\begin{abstract}
We discuss all possible fermion field renormalization prescriptions in conventional field renormalization meaning and mainly pay attention to the imaginary part of unstable fermion Field Renormalization Constants (FRC). We find that introducing the off-diagonal fermion FRC leads to the decay widths of physical processes $t\rightarrow c\,Z$ and $b\rightarrow s\,\gamma$ gauge-parameter dependent. We also discuss the necessity of renormalizing the bare fields in conventional quantum field theory.
\end{abstract}

\maketitle

Generally people think the bare fields need to be renormalized and the introduced FRC is equivalent to the Wave-function Renormalization Constant (WRC). Although the FRC is present for a long time, its imaginary part for unstable particles hasn't been exactly defined \cite{c1,c2}. Since the FRC can be present in physical amplitudes, its imaginary part may affect the physical results. Thus it is very necessary to investigate the definition of the imaginary part of unstable particles' FRC. Here we mainly discuss how to introduce the off-diagonal fermion FRC and investigate the gauge dependence of the physical results induced by the off-diagonal fermion FRC. Besides, we discuss the necessity of renormalizing the bare fields in conventional quantum field theory.

The fermion FRC containing off-diagonal part can be introduced as \cite{c2}
\beq
  \Psi_{0i}\,=\,\sum_j Z^{\half}_{ij}\Psi_j\,, \hspace{10mm}
  \bar{\Psi}_{0i}\,=\,\sum_j \bar{\Psi}_j\bar{Z}^{\half}_{ji}\,,
\eeq
with the fermion FRC 
\beq 
  Z^{\half}_{ij}\,=\,Z^{L\half}_{ij}\gamma_L + Z^{R\half}_{ij}\gamma_R\,,\hspace{10mm}
  \bar{Z}^{\half}_{ij}\,=\,\bar{Z}^{L\half}_{ij}\gamma_{R} +
  \bar{Z}^{R\half}_{ij}\gamma_{L}\,, 
\eeq
where $\gamma_L$ and $\gamma_R$ are the left- and right- handed helicity operators. The fermion FRC $Z^{1/2}_{ij}$ and $\bar{Z}^{1/2}_{ij}$ must satisfy the `pseudo-hermiticity' relationship \cite{c3,c2}
\beq
  \bar{Z}^{\half}_{ij}\,=\,\gamma^{0}Z^{\half\dagger}_{ij}\gamma^{0}\,,
\eeq
since the bare fermion fields must satisfy the condition $\bar{\Psi}_{0i}=\Psi_{0i}^{\dagger}\gamma^0$. 

Under the requirement of Eq.(3) we investigate what field renormalization prescription is acceptable. Within one-loop accuracy one has $Z^{1/2}_{ij}=1+\delta Z_{ij}/2$ and $\bar{Z}^{1/2}_{ij}=1+\delta\bar{Z}_{ij}/2$. Thus the renormalized fermion one-loop self-energy function can be written as
\beqa
  \begin{picture}(72,16)
      \SetOffset(0,-3)
      \ArrowLine(0,5)(24,5)
      \GCirc(36,5){12}{0.5}
      \ArrowLine(48,5)(72,5)
      \Text(10,13)[]{$j$}
      \Text(10,-3)[]{$p$}
      \Text(62,13)[]{$i$}
  \end{picture}\,=\,i\hat{\Gamma}_{ij}(p)\,=&&\hspace{-3mm}\,i(\xslash p-m_i)\delta_{ij}
  +i\left[ {\xslash p}\gamma_L\Sigma^L_{ij}(p^2)+{\xslash p}\gamma_R\Sigma^R_{ij}(p^2)
  +(m_i\gamma_L+m_j\gamma_R)\Sigma^S_{ij}(p^2) \right. \nonumber \\
  &&\hspace{-3mm}+ \left.\hspace{-1mm} \delta\bar{Z}_{ij}(\xslash p-m_j)/2+
  (\xslash p-m_i)\delta Z_{ij}/2-\delta m_i\delta_{ij} \right]\,.
\eeqa
If we introduce the conventional off-diagonal fermion field renormalization conditions
\beq
  \hat{\Gamma}_{ij}(p)\,u_j(p)|_{p^2=m_j^2}\,=\,0\,, \hspace{10mm}
  \bar{u}_{i}(p)\hat{\Gamma}_{ij}(p)|_{p^2=m_i^2}\,=\,0\,, \hspace{10mm} i\not=j\,,
\eeq
we will encounter a bad thing that there is no solution for Eq.(5) if we keep the `pseudo-hermiticity' relationship of Eq.(3), because the existence of the imaginary parts which come from the loop momentum integrals makes the fermion self-energy functions $\Sigma^L$, $\Sigma^R$ and $\Sigma^S$ un-Hermitian \cite{c2}. The only off-diagonal fermion field renormalization conditions which satisfy Eq.(3) are \cite{c4}
\beq
  \tilde{Re}\,\hat{\Gamma}_{ij}(p)\,u_j(p)|_{p^2=m_j^2}\,=\,0\,, \hspace{10mm}
  \tilde{Re}\,\bar{u}_{i}(p)\hat{\Gamma}_{ij}(p)|_{p^2=m_i^2}\,=\,0\,,
  \hspace{10mm} i\not=j\,,
\eeq
where $\tilde{Re}$ takes the real part of the loop momentum integrals appearing in the self energies but not of the coupling constants appearing there which is equivalent to the {\em quasi}-real part defined in Ref.\cite{c5}, because after removing the imaginary part of the loop momentum integrals the fermion self-energy functions $\Sigma^L$, $\Sigma^R$ and $\Sigma^S$ become Hermitian. The corresponding solutions of Eq.(6) are \cite{c4}
\beqa  
  \delta Z_{ij}^{L}&=&\frac{2}{m_i^2-m_j^2}\tilde{Re}\left[ m_j^2 \Sigma_{ij}^L(m_j^2)
  +m_i m_j \Sigma_{ij}^R(m_j^2)+(m_i^2+m_j^2)\Sigma_{ij}^{S}(m_j^2) \right]\,,
  \hspace{6mm}i\not=j\,, \nonumber \\
  \delta Z_{ij}^{R}&=&\frac{2}{m_i^2-m_j^2}\tilde{Re}\left[ m_i m_j \Sigma_{ij}^L(m_j^2)
  +m_j^2 \Sigma_{ij}^R(m_j^2)+2\,m_i m_j \Sigma_{ij}^{S}(m_j^2) \right]\,,
  \hspace{12mm}i\not=j\,.
\eeqa

But such fermion field renormalization prescription leads to the physical amplitudes gauge-parameter dependent \cite{c2}. Furthermore our calculations show that such fermion field renormalization prescription leads to the physical results gauge-parameter dependent. Consider the physical process of top quark decaying into charm quark and gauge boson Z one has to one-loop level
\beq
  {\cal M}(t\rightarrow c\,Z)\,=\,\frac{e(4 s_W^2-3)}{12 s_W c_W}
  (\delta Z^L_{ct}+\delta\bar{Z}^L_{ct})\bar{c}\,{\xslash \epsilon^{\ast}}\gamma_L\,t+
  \frac{e\,s_W}{3 c_W}(\delta Z^R_{ct}+\delta\bar{Z}^R_{ct})\bar{c}\,{\xslash \epsilon^{\ast}}\gamma_R\,t+
  {\cal M}^{amp}(t\rightarrow c\,Z)\,,
\eeq
where $e$ is electron charge, $s_W$ and $c_W$ are the sine and cosine of the weak mixing angle, and ${\cal M}^{amp}$ is the amplitude of the one-loop amputated diagrams shown in Fig.1.
\begin{figure}[htbp]
\begin{center}
  \epsfig{file=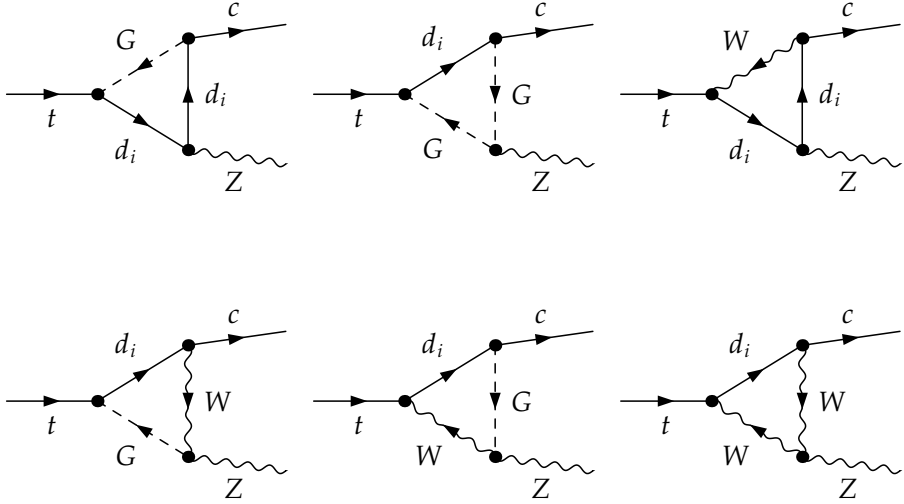,width=12cm} \\
  \caption{One-loop amputated diagrams of $t\rightarrow c\,Z$.}
\end{center}
\end{figure}
The numerical results have shown that the {\em quasi}-real part of ${\cal M}(t\rightarrow c\,Z)$ is gauge-parameter independent \cite{c5}
\beq
  \tilde{Re}{\cal M}(t\rightarrow c\,Z)|_{\xi}\,=\,0\,,
\eeq
where the subscript $\xi$ represents the gauge-parameter dependent part. But the {\em quasi}-imaginary part, which takes the imaginary part of the loop momentum integrals appearing in the amplitude but not of the coupling constants appearing there, of ${\cal M}(t\rightarrow c\,Z)$, isn't the case. Since there is no {\em quasi}-imaginary part in the quark FRC under the conditions of Eq.(7), we only need to calculate the {\em quasi}-imaginary part of ${\cal M}^{amp}(t\rightarrow c\,Z)$. After careful calculations we obtain by the {\em cutting rules} \cite{c6,c5}
\beqa
  \tilde{Im}{\cal M}(t\rightarrow c\,Z)|_{\xi}\,=\hspace{-3mm}&&\bar{c}\,
  {\xslash \epsilon^{\ast}}\gamma_L\,t\,\sum_i\frac{V_{2i}V^{\ast}_{3i}\,e^3
  (4 s_W^2-3)}{384\pi\,c_W\,s_W^3}\,\Bigl [ \nonumber \\
  &&\frac{x_c-\xi_W-x_{d,i}}{x_c}\sqrt{x_c^2-2(\xi_W+x_{d,i})x_c+(\xi_W-x_{d,i})^2}\,
  \theta[m_c-m_{d,i}-M_W\sqrt{\xi_W}] \nonumber \\
  +\hspace{-3mm}&&\left. \frac{x_t-\xi_W-x_{d,i}}{x_t}
  \sqrt{x_t^2-2(\xi_W+x_{d,i})x_t+(\xi_W-x_{d,i})^2}\,
  \theta[m_t-m_{d,i}-M_W\sqrt{\xi_W}] \right],
\eeqa
where $\theta$ is the Heaviside function, $V_{2i}$ and$V_{3i}$ are CKM matrix elements \cite{c7}, $m_c$, $m_t$ and $m_{d,i}$ are the masses of charm quark, top quark and down-type $i$ quark, $M_W$ and $\xi_W$ are the mass and gauge parameter of the gauge boson $W$, and $x_c=m_c^2/M_W^2$, $x_t=m_t^2/M_W^2$, $x_{d,i}=m^2_{d,i}/M^2_W$. We note that the result of Eq.(10) coincides with the results of the conventional loop momentum integral algorithm (see Eqs.(8,9) of Ref.\cite{c5}) and the causal perturbative theory \cite{c8}. Since there is no tree level contribution, Eqs.(9,10) means the decay width of $t\rightarrow c\,Z$ is gauge-parameter dependent under the fermion field renormalization prescription of Ref.\cite{c4}. In order to throw off any suspicion we list the result of $\tilde{Im}{\cal M}(t\rightarrow c\,Z)$ obtained by the optical theorem in appendix, which also gets to the same conclusion. 

Another similar example is the calculation of the decay width of bottom quark decaying into strange quark and photon. To one-loop level one has
\beq
  {\cal M}(b\rightarrow s\,\gamma)\,=\,-\frac{e}{6}(\delta Z^L_{sb}+\delta\bar{Z}^L_{sb})
  \bar{s}\,{\xslash \epsilon^{\ast}}\gamma_L\,b-\frac{e}{6}(\delta Z^R_{sb}+
  \delta\bar{Z}^R_{sb})\bar{s}\,{\xslash \epsilon^{\ast}}\gamma_R\,b+
  {\cal M}^{amp}(b\rightarrow s\,\gamma)\,,
\eeq
where ${\cal M}^{amp}(b\rightarrow s\,\gamma)$ is the amplitude of the one-loop amputated diagrams shown in Fig.2.
\begin{figure}[htbp]
\begin{center}
  \epsfig{file=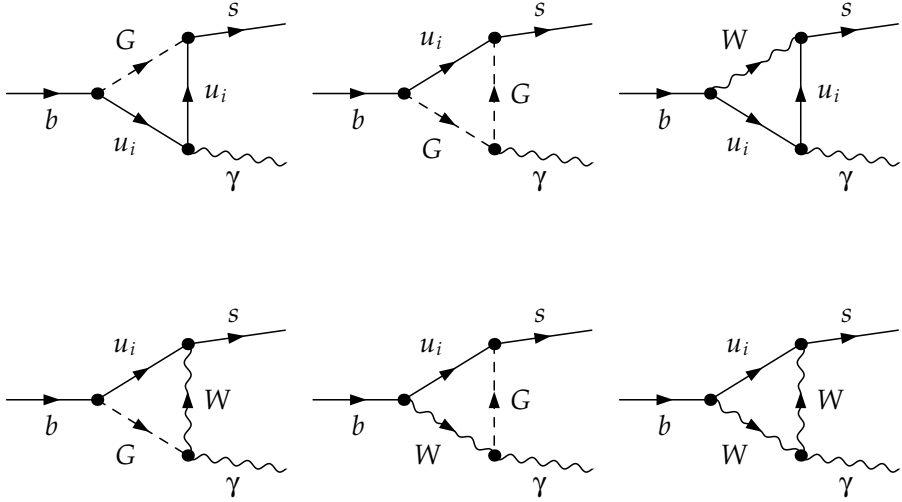,width=12cm} \\
  \caption{One-loop amputated diagrams of $b\rightarrow s\,\gamma$.}
\end{center}
\end{figure}
The {\em quasi}-real of ${\cal M}(b\rightarrow s\,\gamma)$ is gauge-parameter independent \cite{c5}
\beq
  \tilde{Re}{\cal M}(b\rightarrow s\,\gamma)|_{\xi}\,=\,0\,.
\eeq
There is no {\em quasi}-imaginary part in ${\cal M}(b\rightarrow s\,\gamma)$ \cite{c5}, but under the conditions of Eqs.(7) we obtain by the {\em cutting rules}
\beqa
  \tilde{Im}{\cal M}(b\rightarrow s\,\gamma)\,=\hspace{-3mm}&&-\bar{s}\,
  {\xslash \epsilon^{\ast}}\gamma_L\,b\,\sum_i\frac{V_{i3}V^{\ast}_{i2}\,e^3}
  {192\pi\,s_W^2}\,\Bigl [ \nonumber \\
  &&\frac{x_s-\xi_W-x_{u,i}}{x_s}\sqrt{x_s^2-2(\xi_W+x_{u,i})x_s+(\xi_W-x_{u,i})^2}\,
  \theta[m_s-m_{u,i}-M_W\sqrt{\xi_W}] \nonumber \\
  +\hspace{-3mm}&&\left. \frac{x_b-\xi_W-x_{u,i}}{x_b}
  \sqrt{x_b^2-2(\xi_W+x_{u,i})x_b+(\xi_W-x_{u,i})^2}\,
  \theta[m_b-m_{u,i}-M_W\sqrt{\xi_W}] \right] \,,
\eeqa
where $m_s$, $m_b$ and $m_{u,i}$ are the masses of strange quark, bottom quark and up-type $i$ quark, and $x_s=m_s^2/M_W^2$, $x_b=m_b^2/M_W^2$, $x_{u,i}=m^2_{u,i}/M^2_W$. Obviously this result also means the fermion field renormalization prescription of Ref.\cite{c4} leads to the physical result gauge-parameter dependent. On the other hand, the result of $\tilde{Im}{\cal M}(b\rightarrow s\,\gamma)$ obtained by the optical theorem (listed in the appendix) also leads to the same conclusion.

Now we encounter a serious problem that the only acceptable fermion field renormalization prescription which contains off-diagonal fermion FRC is un-acceptable. The main ruination is the constraint of Eq.(3) which is broken by the singularities of Feynman amplitudes \cite{c5}. One way to solve this problem is to introduce two set FRC $Z$ and $\bar{Z}$, $Z$ is for the incoming fermion fields and out-going anti-fermion fields, $\bar{Z}$ is for the out-going fermion fields and incoming anti-fermion fields \cite{c1,c2}. $Z$ and $\bar{Z}$ are introduced as Eqs.(1), but they haven't any relationship between each other. Such prescriptions can make the decay widths of $t\rightarrow c\,Z$ and $b\rightarrow s\,\gamma$ gauge-parameter independent \cite{c5}, but $Z$ and $\bar{Z}$ aren't the FRC in conventional meaning because they doesn't satisfy the necessary field renormalization conditions of Eq.(3). In other words, such prescriptions shouldn't be classified as field renormalization prescription. 

The other way to solve this problem is not to introduce the off-diagonal fermion FRC. The diagonal fermion field renormalization conditions can be introduced as follows
\beqa
  \hat{\Gamma}_{ii}(p)\,u_i(p)|_{p^2=m_i^2}\,=\,0\hspace{-3mm}&&, \hspace{10mm}
  \bar{u}_{i}(p)\hat{\Gamma}_{ii}(p)|_{p^2=m_i^2}\,=\,0\,, \nonumber \\
  \lim_{p^2\rightarrow m_i^2}\frac{{\xslash p}+m_i}{p^2-m_i^2}Re\,\hat{\Gamma}_{ii}(p)\,
  u_i(p)\,=\,u_i(p)\hspace{-3mm}&&, \hspace{10mm}
  \lim_{p^2\rightarrow m_i^2}\bar{u}_i(p)Re\,\hat{\Gamma}_{ii}(p)
  \frac{{\xslash p}+m_i}{p^2-m_i^2}\,=\,\bar{u}_i(p)\,.
\eeqa
Note that the on-shell mass renormalization prescription has been used in Eq.(14). But such renormalization prescription discard the gauge-dependent imaginary part of self-energy functions thus will lead to the physical amplitudes gauge-parameter dependent \cite{c2}. 

In fact, we don't need to renormalize the bare fields in conventional quantum field theory. In a quantum field theory the Hamiltonian can be divided into freedom part and interaction part: 
\beq
  H\,=\,H_0+H_{int}\,.
\eeq
The {\em interaction picture} field is defined as a free field: 
\beq
  \frac{\partial \phi_I (x)}{\partial t}\,=\, i [H_0, \phi_I ]\,.
\eeq
The Heisenberg field which contains interactions can be expressed in terms of $\phi_I$: 
\beq
  \phi_H(t,{\bf x})\,=\, U^{\dagger}(t,t_0)\phi_I(t,{\bf x})U(t,t_0)\,, 
\eeq
with the unitary matrix
\beq
  U(t,t_0)\,=\,T\left\{exp[-i\int_{t_0}^t d t^{\prime} H_{int}(t^{\prime})]\right\} \,,
\eeq
where $T$ is the time-ordering operator and $t_0$ is a reference time. Since $\phi_I$ doesn't contain interactions, it doesn't need to be renormalized. Only the Heisenberg field needs to be renormalized. But it is well known that the conventional Hamiltonian is expressed in terms of $\phi_I$ and the conventional perturbative expansion prescription (i.e. Feynman diagram expansion prescription) uses $\phi_I$ to calculate the S-matrix elements. So it doesn't need to renormalize the bare fields in conventional quantum field theory. On the other hand, according to the LSZ reduction formula the WRC, i.e. the field strength renormalization factor \cite{c9}:
\beq
  Z^{\half}\,=\,<\Omega\,|\,\phi_H(0)\,|\,\lambda_0>\,,
\eeq
with $\Omega$ the interaction vacuum and $\lambda_0$ the in-{\small /}out- particle state of S-matrix elements with zero space momentum, which is a Heisenberg field expectation between the interaction vacuum and the in-{\small /}out- particle state of S-matrix elements, thus one can obtain the WRC from the corrections to the external-line particles of S-matrix elements. Doing so implies the interaction picture field is equivalent to the in-{\small /}out- particle state of S-matrix elements. This is an acceptable hypothesis because the in-{\small /}out- particle state in physical experiments is far away from the interaction region thus is an asymptotic free particle state \cite{c10}.

In summary, we have shown that, to one-loop order, the only acceptable fermion field renormalization prescription which contains off-diagonal fermion FRC leads to the decay widths of $t\rightarrow c\,Z$ and $b\rightarrow s\,\gamma$ gauge-parameter dependent. We advocate not to renormalize the bare fields in conventional quantum field theory because the bare fields conventionally used in Hamiltonian and perturbative calculations of S-matrix elements are free fields not containing interactions. The WRC can be obtained from the corrections to the external-line particles of S-matrix elements according to LSZ reduction formula. This problem will be investigated in future.

We note that our conclusion is also suitable for boson and can be tenable for any other quantum field theory beyond SM.

\vspace{5mm} {\bf \Large Acknowledgments} \vspace{2mm} 

The author thanks professor Xiao-Yuan Li for his useful guidance and Dr. Yu-qi Li for the fruitful discussions with him.

\vspace{5mm} {\bf \Large Appendix} \vspace{2mm}

Because their are some differences between the {\em cutting rules} and the optical theorem \cite{c5}, we list the results of $\tilde{Im}{\cal M}(t\rightarrow c\,Z)$ and $\tilde{Im}{\cal M}(b\rightarrow s\,\gamma)$ obtained by the optical theorem in this appendix. Under the fermion field renormalization prescription of Ref.\cite{c4} we obtain
\beqa
  \tilde{Im}{\cal M}(t\rightarrow c\,Z)^{op}|_{\xi}\,=\hspace{-3mm}&&\bar{c}\,
  {\xslash \epsilon^{\ast}}\gamma_L\,t\,\sum_i\frac{V_{2i}V^{\ast}_{3i}\,e^3
  (4 s_W^2-3)(x_t-\xi_W-x_{d,i})}{384\pi\,c_W\,s_W^3 x_t} \nonumber \\
  &&\times \sqrt{x_t^2-2(\xi_W+x_{d,i})x_t+(\xi_W-x_{d,i})^2}\,
  \theta[m_t-m_{d,i}-M_W\sqrt{\xi_W}]\,,
\eeqa
where the superscript $op$ denotes the result is obtained by the optical theorem. Obviously Eq.(20) is also gauge-parameter dependent. Similarly, for the process of $b\rightarrow s\,\gamma$ we obtain under the fermion field renormalization prescription of Ref.\cite{c4} 
\beqa
  \tilde{Im}{\cal M}(b\rightarrow s\,\gamma)^{op}\,=\hspace{-3mm}&&-\bar{s}\,
  {\xslash \epsilon^{\ast}}\gamma_L\,b\,\sum_i\frac{V_{i3}V^{\ast}_{i2}\,e^3
  (x_b-\xi_W-x_{u,i})}{192\pi\,s_W^2 x_b} \nonumber \\
  &&\times \sqrt{x_b^2-2(\xi_W+x_{u,i})x_b+(\xi_W-x_{u,i})^2}\,
  \theta[m_b-m_{u,i}-M_W\sqrt{\xi_W}]\,.
\eeqa
Eq.(21) is also gauge-parameter dependent.

\end{document}